\documentclass[a4paper,11pt]{article}
\usepackage{pos}

\title{Acoustic detection of UHE neutrinos: ANDIAMO perspectives}

\author*[a,b,c]{Antonio Marinelli}
\author[b]{Pasquale Migliozzi}
\author[b]{Andreino Simonelli}

\affiliation[a]{Dipartimento di Fisica ``Ettore Pancini''  , Universit\'a degli studi di Napoli ``Federico II'' , \\
  Complesso Univ. Monte S. Angelo, I-80126, Napoli, Italy}

\affiliation[b]{INFN - Sezione di Napoli, \\ Complesso Univ. Monte S. Angelo, I-80126 Napoli, Italy}
\affiliation[c]{INAF-Osservatorio Astronomico di Capodimonte, \\ Salita Moiariello 16, I-80131 Naples, Italy}

\emailAdd{antonio.marinelli@unina.it}
\emailAdd{pasquale.migliozzi@na.infn.it}
\emailAdd{andreino.simonelli@na.infn.it}

\abstract{
A possible detection of ultra-high-energy neutrinos has been attempted since decades through the Askarian radiation and different observation techniques. In fact, when such energetic neutrinos interact in a medium are able to produce a thermo-acoustic effect resulting in a bipolar pressure pulse that carries a portion of the energy generated by the particle cascades. This effect can be observed in atmosphere looking for the correlated radio emission and in ice/water searching directly the acoustic pulse. The kilometric attenuation length as well as the well-defined shape of the expected pulse favors a large-area-undersea-array of acoustic sensors as a possible observatory. Previous efforts of taking data with a undersea hydrophones array were obtained thanks to already installed submarine military arrays or acoustic system built to calibrate the positions of Cherenkov light detector units.  In this proceeding we propose to use the based but not operative offshore oil rigs powered platforms in the Adriatic sea as the main infrastructure to build an acoustic submarine array of dedicated hydrophones covering a total surface area up to $\sim$10000 Km$^{2}$ and a volume up to $\sim$500 Km$^{3}$. A future identification of neutrino events at energies greater than 10$^{18}$ eV will confirm the presence of powerful accelerators in our Universe able to emit cosmic rays up to ZeV energy range.}

\FullConference{%
   27th European Cosmic Ray Symposium - ECRS \\
   25-29 July 2022 \\
   Nijmegen, the Netherlands 
}

\begin{document}
\maketitle

\section{Introduction}
The ultra-high-energy cosmic-rays (UHECRs) collected by Pierre Auger experiment~\cite{Abraham:2010mj} and Telescope Array~\cite{AbuZayyad:2012ru} confirm the presence of astrophysical accelerators who can originate EeV neutrinos when such energetic charged particles interact with gas or photons inside the astrophysical sources or in the intergalactic medium~\citep{Bahcall:1999yr}. The multi-messenger description of IceCube observations~\cite{Aartsen:2014gkd} and Fermi-LAT telescope suggest the possibility that several extragalactic accelerators can peak their neutrino emission in the range of energies between several hundreds of TeV up to a few EeVs. The hadronic production in this energy range can be made by different class of AGNs as well as by powerful GRBs~\citep{Murase:2005hy}, however we don't have at the moment measurements that can confirm these hypotheses. Moreover in addition to the neutrino directly produced at the source, in this energy range, we can expect also a flux of neutrino produced by the interaction of ultra-high-energy cosmic rays (UHECRs)  with the astrophysical background radiation fields, the so called cosmogenic neutrinos. This photon target for the production of these neutrinos is represented by the extra-galactic background light (EBL) and the cosmic microwave backgrounds (CMB) along the path between the source and the Earth~\citep{Beresinsky:1969qj}. While the former neutrino flux can produce a spotted observation in the skymap, possibly correlated with UHE sources, the latter is expected to be uniform over the full sky. Spectral energy distributions (SEDs) of neutrinos above the PeV energy will also be crucial to constrain the different extragalactic components verifying the hypotheses made to explain the UHE cosmic rays extending at higher energies the observations of IceCube~\citep{IceCube:2016zyt}, Antares~\citep{ANTARES:2011hfw}, Baikal~\citep{Aynutdinov:2012zz} and KM3NeT~\citep{KM3Net:2016zxf}. \\ The two major techniques proposed to observed UHE neutrinos are the radio detection in atmosphere and  the acoustic detection in water/ice thanks to the Askarian effect~\citep{Askaryan:1962hbi} happening when high-energy ionizing particles pass through a dense medium. The two experimental setups represents a complementary approach covering a bigger portion of the sky, while the first favours the Earth-skimming neutrino events, the second privileges the down-going events.
The radio detection techniques improved during the last decades thanks to the installation of several prototypes and telescopes as the pioneering RICE~\cite{Kravchenko:2001id,Kravchenko:2011im}, ARA~\cite{Allison:2015eky} , ARIANNA~\cite{Barwick:2014rca,Barwick:2014pca}, RNO-G~\cite{Aguilar:2020xnc}, GRAND~\cite{2020SCPMA..6319501A} as well as the balloon experiments like ANITA~\cite{Gorham:2008dv}. \\
Instead for the acoustic detection, only small acoustic hydrophone arrays were used to take data until now, like SPATS~\cite{Abdou:2011cy}, O$\nu$DE5\cite{Riccobene:2009zz}, ACoRNE~\cite{Danaher:2007zz}, AMADEUS~\cite{Aguilar:2010ac} and SOUND~\cite{Vandenbroucke:2004gv}, being built for sea monitoring activities or constructed as sub-detector of major Cherenkov telescopes. To finalize the construction of a dedicated acoustic array we should consider the coverage of a large area $\mathcal{O}(1000~$Km$^{2})$ and a correspondent sizeable volume $\mathcal{O}(100~$Km$^{3})$~\cite{Kurahashi:2010ei} of instrumented water.\\
In this proceeding we report the idea of exploiting the ENI not operative powered oil rigs in the offshore of the Adriatic sea. The oil upstream activity is terminated since few years, however the infrastructures are still available for other purposes, like scientific activities. With this proposed telescope, ANDIAMO (Acoustic Neutrino Detection in a Multidisciplinary Observatory), we aim to build the biggest undersea acoustic array built up to date, taking advantages from the Mediterranean favorable properties like the average water temperature, currents, and salinity. Here we show the expected propagation of the acoustic signal produced in a shallow water environment and we report the possible sensitivity, demonstrating that a interesting discovery potential can be reached within a decade of ANDIAMO data taking.

\subsection{The existing infrastructure in Adriatic Sea} 
The platforms that can be used for the ANDIAMO experiment are positioned in the offshore of Adriatic Sea at a distance from the shore ranging from 10 to 60 Km, following roughly the natural shape of the continental margin where oil and gas were extracted. The depth reached for these installations, is typical of a continental shelf, with depth ranging from 25 to 80 meters. The distribution of platforms is reported in Fig~\ref{zerodegmap} with the corresponding sensitive area for a vertical UHE neutrino event.\\
The shallow depth of the ENI platforms provide us several advantages, with the most important one in terms of geometrical spreading. The acoustic wave generated by neutrino interaction never goes in the far field regime where spherical spreading is characteristic, this is the case for acoustic detectors implemented on optical arrays for example, where the array dimensions are small compared to the extension of the channel where they are installed. In our case, the cascade length is comparable to the depth of the channel, therefore we do not expect any transition region from near to far field propagation. The amplitude of the acoustic wave will decay as $1/\sqrt r$ respect to a spherical spreading where the decay is $1/r$. In terms of attenuation, this environment has 10 dB/decade against 20 dB/decade for spherical spreading. In simple words, we really deal with a cylindrical wave instead of a spherical one.
Moreover, the relatively high temperature of the shallow Adriatic Sea together with a lower salinity due to the fresh water influx from the largest Italian river, the Po river, will contribute to decrease the attenuation too. With these peculiarities the Mediterranean Sea can be considered one of the best environments for a underwater neutrino telescope as already underlined by~\citep{Niess2006}. In addition, also the logistic will play in favor of this solution since power and connectivity are supposed to be already present in the ENI oil rigs. Moreover the steady structure of the platforms will permit to install a mini array of hydrophones in every structure with a solution similar to the one adopted in~\citep{Aynutdinov:2012zz}.\\
To carefully address the gains of this solution the underwater noise should be studied with dedicated data taking campaigns. The stationary noise floor for our system is mostly dependent on sea state in the frequency range of detection for the predicted acoustic signal (1-20 kHz).
Transient noises as dolphins and propellers can of course spoil the sensitivity locally and in an unpredictable manner but in a long term perspective can be filtered and recognized since they have a different time/frequency signature compared to the neutrino generate ones. Sound propagation in very shallow waters has been studied largely in literature by~\citep{Kuperman2004} with a common conclusion that shallow water propagation can be seen as a small scale problem respect to deep waters; considering that we applied ray tracing to simulate our environment. The structure of the velocity profile resulted to be a game changer in the possibility to design an effective acoustic telescope, in particular here we want to exploit the presence of the SOFAR channel to stretch the acoustic detection range to the maximum. As we shown~\citep{Marinelli:2021upw}, with more details reported, the available geometry of platforms distribution seems suitable for the realization of the largest acoustic array for UHE neutrinos detection never built, with a sensitive area of more than 10000 Km$^{2}$.

\section{Generated acoustic signal and reconstruction technique}
\begin{figure}
\includegraphics[width=0.6\textwidth]{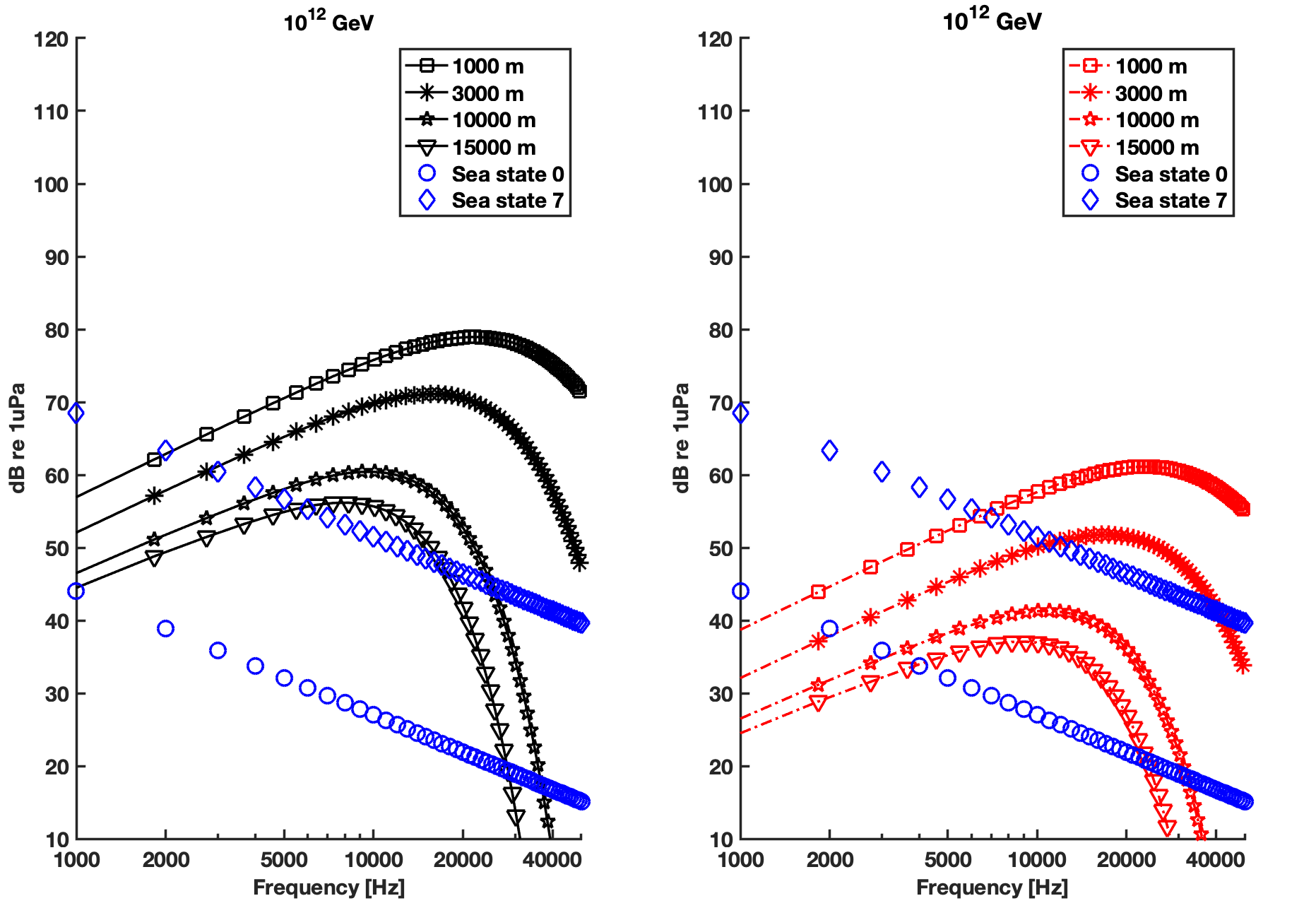}
\centering
\caption{Calculated spectra for different sea depths with fixed shower energy, on the left panel a 30 meters deep sea and on the right panel a 3000 meters deep sea, we assume iso-velocity conditions.}
\label{cfrdepth}
\end{figure}

\begin{figure}[h!]
    \centering
    \includegraphics[width=0.5\textwidth]{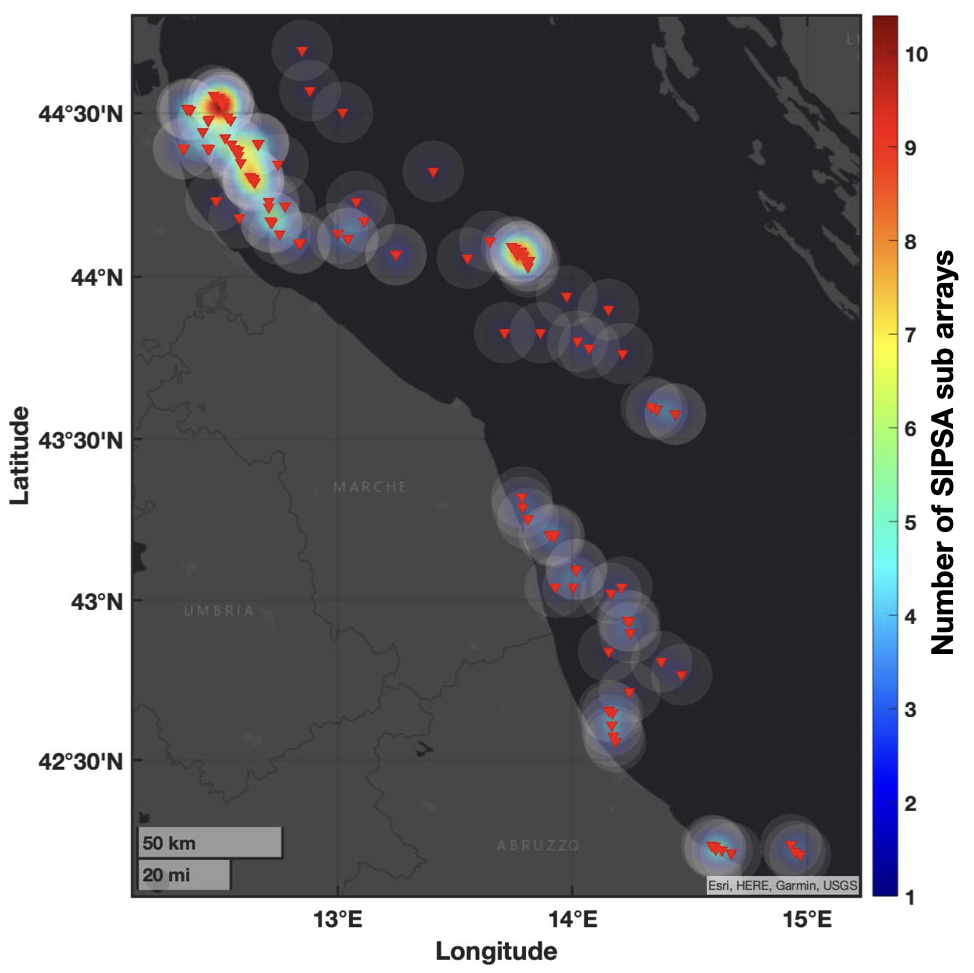}
    \caption{Map illustrating the covered sensitive area for vertical neutrino induced showers at $10^{19}$ eV, the color code represent the density of sub arrays SIPSA that can potentially detect a signal generated inside the covered area.}
    \label{zerodegmap}
\end{figure}

When a UHE neutrino interacts with matter can decay with different respective channels and generate a hadronic cascade. The deposition of the energy released by the cascade in a small volume of water  takes a cylindrical shape with a length of few tens of meters and a few centimeters in radius, heats the medium and expand it. The quasi-instantaneous expansion of the water results in a coherent production of a collimated mechanical wave propagating perpendicularly to the shower axis. The extension in depth of the cascade gives the length of the cylinder where the thermo-acoustic effect occurs. The intensity of this cylindrical wave expansion, also called  pancake, depends on the angle of the cascade with respect to the vertical and the decay channel that produced the cascade. The waveform generated by a UHE neutrino interaction resembles a bipolar signal, whose amplitude at 100 meters from the cascade production can range from 10 Pascals to some milliPascal. The central frequency is expected to be around 50 kHz at 100 meters with a duration of the order of several microseconds according to simulations proposed by~\citep{BEVAN2009398}.
The pulse propagation is affected by the attenuation related to geometrical spreading and absorption, with a consequent broadening of the wavelet in time and an amplitude decrease.
However modern hydrophones have a self-noise level that is lower than the sea state level zero in the Wenz sea level scale \citep{wenz}. Therefore, using the most sensitive hydrophones, we are left with the problem of a realistic calculation of the attenuation and frequency content change in the signal versus distance. In~\citep{Marinelli:2021upw} it is reported in detail the expected amplitude spectrum as well as the attenuated spectra for different source levels for different distances from the neutrino event and different frequency of the signal. Here we report, for a fixed energy, the dispersion of the spectrum for a 3000 meters deep channel in comparison to the one for a 30 meters deep channel, to highlight the advantage of shallow waters (see Fig.~\ref{cfrdepth}).
From the spectra, it is evident that above a cascade energy of $10^{19}$ GeV signal can be measured by any single array element (oil rig) that stays in a 10 Km radius from the source event.  
For this calculation we assume a linear path propagation from source to receiver induced by an iso-velocity condition for simplicity. 
Moreover we calculate the transmission losses (TL) for every spectral component of the source signal and using the following equation:
\begin{equation}
    TL=10 \log \frac{I_s}{I(r)}
    \label{tloss}
\end{equation}
we obtain in the end the sound pressure level (SPL) spectra at different distances for different source energies and incidence angle of $0^{\circ}$, the angular dependence of attenuation it is shown in~\citep{Marinelli:2021upw}  where we take into account the seafloor reflection.
The definition of SPL is the usual one for acoustic measurements i.e. $SPL= 20log(P/P_0)$ measured in dB relative to 1 $\mu$Pascal.
The typical shallow water conditions are well studied for the underwater acoustic propagation~\citep{Kuperman2004}; we can summarize the part of our interest with the following scenario that is confirmed by experimental data e.g. for Mediterranean sea. In shallow water conditions we have a first mixed surface layer where the speed of sound is constant, followed by a thermocline and an increase due to pressure as shown in~\citep{Marinelli:2021upw}. \\
In~\citep{Marinelli:2021upw} it is simulated a cylindrical source of 20 meters of extension in depth (cascade length). It is shown that is possible to ray-trace the direction of the wave-fronts generated at different shower angle for a 5 kHz tone which is, according to spectral simulations, the peak level at the distances of interest given the actual platforms geometry. Additional signal attenuation induced by transmission losses at the sea floor are accounted for rays that are not confined inside the SOFAR channel. The transmission losses at the sea floor constitute a severe limiting factor for angles larger than 45 degrees since after this value we have a very steep drop in the reflection coefficient. Instead for smaller angles respect to the zenith, between $\pm 20^{\circ}$ the channeling effect allows the focusing of the energy inside the SOFAR channel and extending the range of the detector, this effect was not achievable in previous arrays like SAUND and ACoRNE. More than this, for a future upgrade, we can think in a denser array pushing further the portion of the sky covered by the ANDIAMO telescope.\\  In~\citep{Marinelli:2021upw} it is proven that the sensitive area of ANDIAMO telescope changes according to angle and energy of the incident neutrino; an example of the covered area is reported in Fig.~\ref{zerodegmap} where the color map represent the SIPSA sub arrays density. In this case for a $10^{19}$ eV cascade vertically originated the sensitive area is of the order of 10000 km$^2$.

\section{Possible sensitivity to UHE neutrinos}
\begin{figure}[h!]
\centering
\includegraphics[scale=0.60]{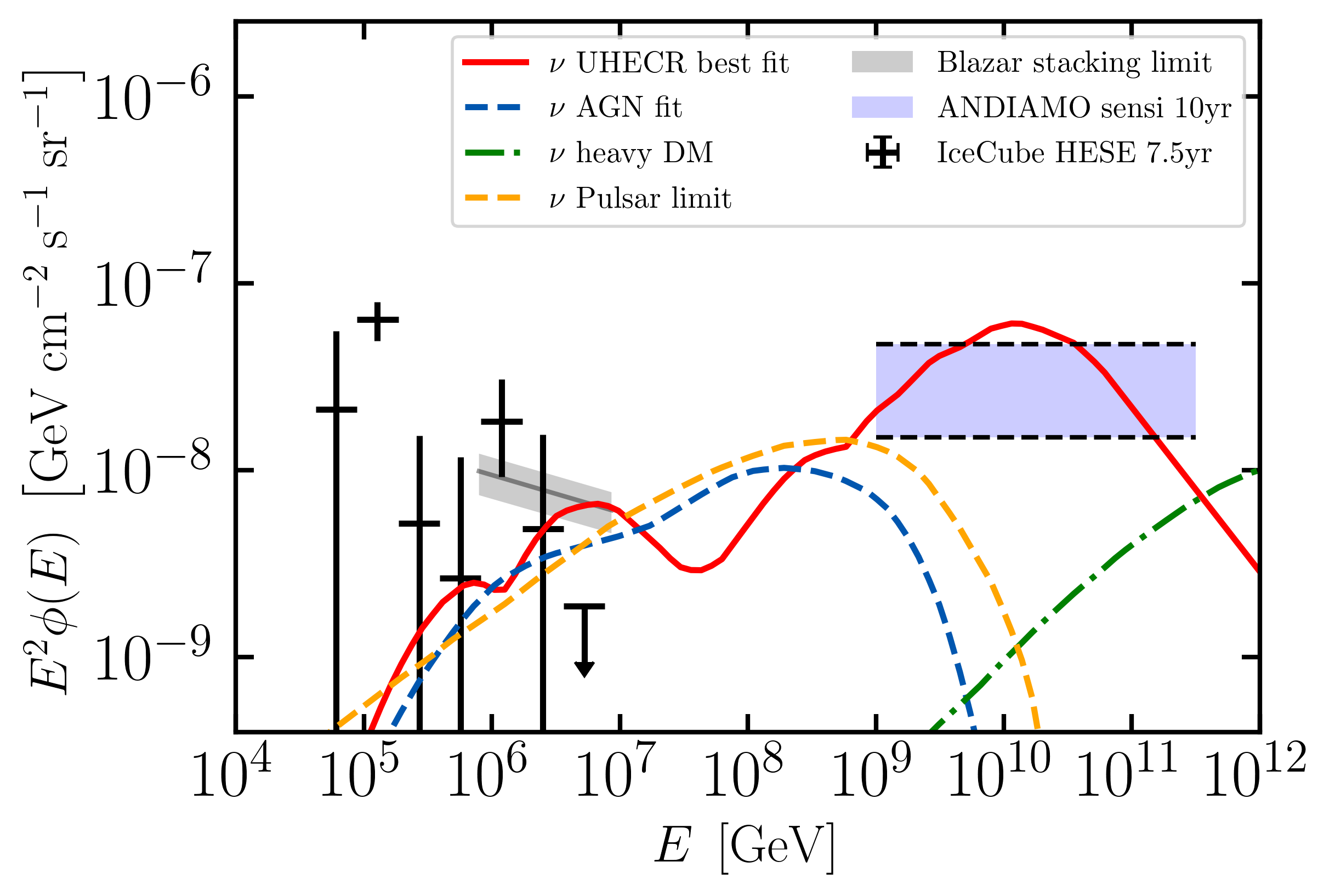}
\caption{In this plot we report the expected sensitivity of the ANDIAMO concept with a surface array equivalent to the one described in Fig.~\ref{zerodegmap} and a observational time of one decade. The three neutrino families are considered with the range of sensitivity spanning the different sensitive areas covered considering the possible beneficial incident angles. For comparison we show also the limit on the possible UHE cosmogenic neutrino flux obtained from the UHECR observations as well as possible UHE neutrino diffuse contributions from AGNs, millisecond pulsars and heavy dark matter decays. IceCube full-sky measurements are reported too.}
\label{ANDIAMO_sensi}
\end{figure}

Even though at moment a precise geometry for the mini-arrays to be installed in each platform is not finalized, in this proceeding as well as in~\citep{Marinelli:2021upw} we obtain preliminary expectations of the ANDIAMO telescope. Being not constructed at the moment a Monte Carlo with the finale geometry, for this scope we can proceed with a semi-analytic approach, as done in~\citep{Marinelli:2021upw} or applying a scaling factor from a previous simulation of a homogeneous hydrophone array with a different instrumented volume. In this proceeding we follow the second approach taking as a reference the simulation reported in~\citep{2017EPJWC.13506002S}. The maximal solid angle around the zenith of the detector is set by the abrupt fall of the reflection coefficient which in turn limits the total distance travelled by the acoustic signal generated by an inclined cascade.
The sensitive area is then angle and energy dependent, for a $10^{19}$ eV cascade it ranges from 10000 km$^2$ for a vertical cascade to about 2800 km$^2$ for a 45$^{\circ}$ inclined one.
This area variability is accounted in the calculation of the sensitivity shown in Fig.~\ref{ANDIAMO_sensi}.
Under the assumption that the UHE neutrino flux behaves like $\Phi(E)=\frac{dN}{dE}=kE^{-2}$ the corresponding sensitivity can be written as:
\begin{equation}
\label{sensi}
k = \frac{N_{evt}}{\int^{E_{max}}_{E_{min}}E^{-2}\mathcal{E}(E)dE},
\end{equation}
where $N_{evt}$ indicates the number of expected events and $\mathcal{E}(E)$ correspond to the exposure obtained through the equation:
\begin{equation}
\label{exposure}
\mathcal{E}(E_{\nu}) = \frac{2\pi}{m}\sum_{i}\left[\sigma_{i}(E_{\nu}) \int dt d\theta dD \sin \theta \cos \theta A^{i}_{eff}(\theta,D,E_{nu},t)\right] ,
\end{equation}
where the sum runs over the three neutrino flavours and the CC and NC interactions, with $\sigma^{i}$ the corresponding $\nu-$nucleon interaction cross-section~\citep{Cooper-Sarkar:2007zsa} and $m$ the nucleon mass. The integral is performed over the zenith angle $\theta$ the interaction depth $D$ of the neutrino (in units of g cm$^{-2}$), and the blind search period. 
As said before in this proceeding we obtain the possible integrated values applying scaling factors to the Monte Carlo calculations reported in~\citep{2017EPJWC.13506002S}.

The expected sensitivity, reported in Fig.~\ref{ANDIAMO_sensi}, is compared with the main expected UHE neutrino diffuse fluxes. Even though more accurate studies can be obtained with the use of a dedicated Monte Carlo simulation chain whenever the geometry will be finalized, the preliminary calculations obtained scaling the sensitivity obtained in~\citep{2017EPJWC.13506002S} for the different volumes of ANDIAMO (with different neutrino incident angles),highlights the potential discoveries of such acoustic array concept.

\section{Conclusions}
The exceptional results reached with VHE neutrino observations in the last decade pointed out the importance of this messenger to better understand the powerful astrophysical accelerators surrounding us. These studies are now lacking of information in the range of UHE, pushing the community to obtain informations about the neutrino spectral features in the energy range from EeV to ZeV. Two main techniques have been explored for the UHE neutrino detection: the extended radio antenna arrays, mainly targeting the eart-skymming neutrino events, and the acoustic underwater/ice hydrophone arrays 
who are more focused on downgoing neutrino events. The latter techniques was carried out in the last decades with small prototypes or through the possible use of hydrophones installed for calibration purposes and not optimized for an acoustic detector design. In this proceeding we report the possibility of building a dedicated acoustic array in the Adriatic Sea using the platforms already installed, and not operative, for oil and gas upstream.\\ The preliminary results, reported more in details in~\citep{Marinelli:2021upw}, seems very promising, showing that a constrain on the expected neutrino spectral features, in the PeV-EeV range, can be possible within a decade of data taking of the ANDIAMO telescope.\\

\bibliography{references}{}
\bibliographystyle{plain}



\end{document}